\documentclass[twocolumn,showpacs,preprintnumbers,amsmath,amssymb]{revtex4}

\usepackage{graphicx}
\usepackage{dcolumn}
\usepackage{bm}

\begin{document}

\title{Decoherence of two qubits in a non-Markovian reservoir without \\ rotating-wave approximation}

\author{Fa-Qiang Wang}
\author{Zhi-Ming Zhang}
\email{zmzhang@scnu.edu.cn}
\author{Rui-Sheng Liang}%
\affiliation{Lab of Photonic Information Technology, School of Information and Photoelectronic Science and Engineering, South China Normal University, Guangzhou 510006, China}%

\date{\today}

\begin{abstract}
The decoherence of two initially entangled qubits in a non-Markovian
reservoir has been investigated exactly without Born Markovian
approximation and rotating-wave approximation(RWA). The
non-perturbative
 quantum master equation is derived and its
 exact solution is obtained. The decoherence behaviors of
 two qubits, initially entangled in Bell states, has been
 investigated in three different cases of parameters. The results show
 that the counter-rotating wave terms have great influence on the
 decoherence behavior, and there are differences between
 the exact solution of the Hamiltonian with RWA
 and that of the exact Hamiltonian without RWA.
\end{abstract}

\maketitle
 \section{Introduction}
\par In recent years, the phenomenon , termed as ``entanglement sudden
death"(ESD), has been found theoretically\cite{yu1,yu2} and shown
experimentally\cite{mp,mf}. It is shown that spontaneous
disentanglement may take only a finite-time to be completed, while
local decoherence (the normal single-atom transverse and
longitudinal decay) takes an infinite time\cite{yu2}. It's quite
different from the case of continuous variable two-atom model
discussed by Dodd and Halliwell\cite{dod}. Thereafter, many works
have been devoted to the related topics\cite{ikr,hu,asm,sch,daj},
and some authors extended the results in Markovian regime to
non-Markovian case\cite{cao,bel}.
\par The RWA, which neglecting counter rotating terms corresponding
to the emission and absorption of virtual photon without energy
conservation, is widely used in quantum optics. Generally, the RWA
is justified for small detunings and small
 ratio of the atom-field coupling divided by the atomic transition
 frequency\cite{kli,Lou,pur}.
 In atom-field  cavity  systems, this ratio is typically of the order  $10^{-7}\sim10^{-6}$.
 Recently, cavity systems with very strong couplings have been discussed\cite{mei}.
 The ratio may also become order of magnitudes larger in solid state systems,
 and the full Hamiltonian, including the  virtual processes
 (counter-rotating terms), must be considered\cite{iri}.
 The neglect of counter-rotating wave terms
in Hamiltonian strongly simplifies the mathematical treatment of the
problem and usually give exact solution of the approximate
Hamiltonian, while the perturbative approach to the systems beyond
the RWA usually is complicated and gives approximate
solution\cite{pur}. Otherwise, with the same parameters, the
approximation in Hamiltonian might lead to different result from the
approximation of exact solution with exact Hamiltonian. Fortunately,
non-perturbative master equation could be obtained by path integral
and it is beneficial to study, exactly, the systems beyond the
RWA\cite{Ish}.
\par In this paper, we will focus on the influence of counter-rotating wave terms on the
decoherence behavoir of two qubits in a non-markovian reservoir from
the exact solution
 of the system beyond the RWA and Born-Markovian approximation.
 In section \ref{sec:model} , the reduced
non-perturbative non-Markovian quantum master equation of an atom in
a non-Markovian reservoir is derived and its the exact solution is
obtained by algebraic approach of Lie superoperator. In section
\ref{sec:num}, the decoherence of two initially entangled atoms
coupled with two cavities separately has been discussed. The
conclusion will be given in section \ref{sec:clu}.
 \section{\label{sec:model} Model and exact solution}
 \subsection{Hamiltonian and non-perturbative master equation}
\par Now we restrict our attention to two noninteracting two-level
atoms A and B coupled individually to two environment
reservoirs\cite{bel}. To this aim, we first consider the Hamiltonian
of the subsystem of a single qubit coupled to its reservoir as
\begin{equation}\label{e1}
    H=H_{a}+H_{r}+H_{ar}
\end{equation}
where
\begin{eqnarray}
  H_{a} &=& \omega_{0}\frac{\sigma_{z}}{2} \\
  H_{r} &=& \sum_{k}\omega_{k}a_{k}^{\dagger}a_{k} \\
  H_{ar} &=& (\sigma_{+}+\sigma_{-})\sum_{k}g_{k}\left(a_{k}^{\dagger}+a_{k}\right)
\end{eqnarray}
where $\omega_{0}$ is the atomic transition frequency between the
ground state $|0\rangle$ and excited state $|1\rangle$.
$\sigma_{z}=|1\rangle\langle1|-|0\rangle\langle0|$,
$\sigma_{+}=|1\rangle\langle0|$ and $\sigma_{-}=|0\rangle\langle1|$
are pseudo-spin operators of atom. The index $k$ labels the field
modes of the reservoir with frequency $\omega_{k}$,
$a_{k}^{\dagger}$ and $a_{k}$ are the modes' creation and
annihilation operators, and $g_{k}$ is the frequency-dependent
coupling constant between the transition $|1\rangle-|0\rangle$ and
the field mode
 $k$.
\par The reduced non-perturbative non-Markovian quantum master equation of atom
 could be obtained by path integral\cite{Ish}
\begin{equation}\label{e5}
    \frac{\partial}{\partial
    t}\rho_{a}=-i\mathfrak{L}_{a}\rho_{a}-\int_{0}^{t}ds\langle \mathfrak{L}_{ar}e^{-i\mathfrak{L}_0(t-s)}\mathfrak{L}_{ar}e^{-i\mathfrak{L}_{0}(s-t)} \rangle_{r}\rho_{a}
\end{equation}
where $\mathfrak{L}_0$, $\mathfrak{L}_a$ and $\mathfrak{L}_{ar}$ are
Liouvillian operators defined as
\begin{eqnarray*}
  \mathfrak{L}_{0}\rho &\equiv& [H_{a}+H_{r},\rho] \\
  \mathfrak{L}_{a}\rho &\equiv& [H_{a},\rho] \\
  \mathfrak{L}_{ar}\rho &\equiv& [H_{ar},\rho]
\end{eqnarray*}
and $\langle...\rangle_{r}$ stands for partial trace of the
reservoir.
\par Then we assume that the reservoir is initially in vacuum states
and the spectral density of the reservoir is in Lorentzian
form\cite{bel,sca}
\begin{eqnarray}\label{ee1}
  J(\omega) &=& \sum_{k}g^{2}_{k}\{\delta(\omega-\omega_{k})+\delta(\omega+\omega_{k})\} \nonumber \\
   &=& \frac{1}{2 \pi} \frac{\lambda \gamma^{2}}{(\omega-\omega_{0})^{2}+\gamma^{2}}
\end{eqnarray}
where $\gamma$ respents the width of the spectral distribution of
the reservoir modes and is related to the correlation time of the
noise induced by the reservoir, $\tau_{r}=1/\gamma$. The parameter
$\lambda$ is related to the subsystem-reservoir  coupling strength.
There are two correlation functions in this model\cite{ccq}
\begin{eqnarray}
  \alpha_{1} &=& \int_{-\infty}^{\infty}J(\omega)e^{-i(\omega-\omega_{0})t}=\frac{\gamma\lambda}{2}e^{-\gamma t} \\
  \alpha_{2} &=& \int_{-\infty}^{\infty}J(\omega)e^{i(\omega+\omega_{0})t}=\frac{\gamma\lambda}{2}e^{(-\gamma+i2\omega_{0} )t}
\end{eqnarray}
$\alpha_{1}$ comes from the rotating-wave interaction and
$\alpha_{2}$ from the counter-rotating wave interaction.
\par So, we could obtain the non-perturbative master equation of the
subsystem from Eq.(\ref{e5})
\begin{eqnarray}\label{e7}
  \frac{\partial}{\partial
    t}\rho_{a} &=& -\frac{\lambda}{2}\left(\gamma\alpha^{R}+f(t)\right)\rho_{a} -i\left(2\omega_{0}-\lambda\gamma \alpha^{I}\right)J_{0}\rho_{a}\nonumber\\
   & & +\frac{\lambda}{2}\left(\gamma\alpha+f(t)\right) J_{+}\rho_{a}+\frac{\lambda}{2}\left(\gamma\alpha^{*}+f(t)\right)
    J_{-}\rho_{a} \nonumber\\
    & & +\lambda \left(\gamma\alpha^{R}-f(t)\right)K_{0}\rho_{a}+\lambda\gamma \alpha^{R}K_{+}\rho_{a} \nonumber\\
    & & + \lambda f(t)K_{-}\rho_{a}
\end{eqnarray}
where $J_{0}$, $J_{+}$, $J_{-}$, $K_{0}$, $K_{+}$ and $K_{-}$ are
superoperators defined as
\begin{eqnarray*}
  J_{0}\rho_{a} &\equiv& \left[\frac{\sigma_{z}}{4},\rho_{a}\right] \\
  J_{+}\rho_{a} &\equiv& \sigma_{+}\rho_{a}\sigma_{+} \\
  J_{-}\rho_{a} &\equiv&  \sigma_{-}\rho_{a}\sigma_{-} \\
  K_{0}\rho_{a} &\equiv& (\sigma_{+}\sigma_{-}\rho_{a}+\rho_{a}\sigma_{+}\sigma_{-}-\rho_{a})/2 \\
  K_{+}\rho_{a} &\equiv& \sigma_{+}\rho_{a}\sigma_{-} \\
  K_{-}\rho_{a} &\equiv& \sigma_{-}\rho_{a}\sigma_{+}
\end{eqnarray*}
and
\begin{equation*}
    \alpha=\frac{1-e^{-(\gamma+i2\omega_{0})t}}{\gamma+i2\omega_{0}}.
\end{equation*}
$\alpha^{R}$, $\alpha^{I}$ and $\alpha^{*}$ are real part, image
part and conjugate of $\alpha$, respectively. $f(t)=1-exp(-\gamma
t)$.

\par From Eq.(\ref{e7}), we find that there is a frequency shift
caused by the interaction between the atom and the non-Markovian
reservoir\cite{breu}. Eq.(\ref{e7}) reveals that the contribution of
counter-rotating terms is in order of
$\lambda\gamma^{2}/\omega_{0}^{2}$ besides the rapidly oscillating
terms, which is quite different from the result of Born and
Markovian approximation of atom in vacuum \cite{agr}.

\subsection{ Exact solution of master equation}
\par The time evolution of density operator in Eq.(\ref{e7}) could be obtained with
algebraic approach in Ref.\cite{zhang} because the superoperators
herein satisfy $SU(2)$ Lie algebraic communication relations, i.e.
\begin{eqnarray}
  \left[J_{-},J_{+}\right]\rho_{a} &=& -2J_{0}\rho_{a} \nonumber\\
  \left[J_{0},J_{\pm}\right]\rho_{a} &=& \pm J_{\pm} \rho_{a} \nonumber\\
  \left[K_{-},K_{+}\right]\rho_{a} &=& -2K_{0}\rho_{a} \nonumber\\
  \left[K_{0},K_{\pm}\right]\rho_{a} &=& \pm K_{\pm}\rho_{a} \nonumber\\
  \left[K_{i},J_{j}\right] &=& 0
\end{eqnarray}
where $i,j=0, \pm$. By directly integrating Eq.(\ref{e7}), the
formal solution is obtained as\cite{pur}
\begin{eqnarray}\label{e8}
  \rho_{a}(t) &=& e^{-\Gamma_{k}}\hat{T}e^{\int_{0}^{t}dt(\varepsilon_{0}J_{0}+ \varepsilon_{+}J_{+}+\varepsilon_{-}J_{-})}\nonumber\\
    & & \times \hat{T}e^{\int_{0}^{t}dt(\nu_{0}K_{0}+\nu_{+}K_{+}+\nu_{-}K_{-})}\rho_{a}(0)
\end{eqnarray}
where $\hat{T}$ is time time order operator,
$\varepsilon_{0}=-i\left(2\omega_{0}-\lambda\gamma \alpha^{I}\right)
$, $\varepsilon_{+}=\lambda\left(\gamma\alpha+f(t)\right)/2$,
$\varepsilon_{-}=\lambda\left(\gamma\alpha^{*}+f(t)\right)/2$,
$\nu_{0}=\lambda\left(\gamma\alpha^{R}-f(t)\right)$, $
\nu_{+}=\lambda\gamma \alpha^{R}$, $\nu_{-}=\lambda f(t)$,
$\Gamma_{k}=\lambda\left(\gamma\tilde{\alpha}^{R}+F(t)\right)/2$,
$F(t)=t-[1-exp(-\gamma t)]/\gamma$ and
\begin{eqnarray}\label{e9}
  \tilde{\alpha} &=& \int_{0}^{t}\alpha dt =\tilde{\alpha}^{R}+i\tilde{\alpha}^{I} \nonumber\\
  \tilde{\alpha}^{*} &=& \tilde{\alpha}^{R}-i\tilde{\alpha}^{I}
\end{eqnarray}
\begin{widetext}
\begin{eqnarray}
  \tilde{\alpha}^{R} &=& \frac{1}{4\omega^{2}_{0}+\gamma^{2}}\left[\gamma
    t+\frac{1}{4\omega^{2}_{0}+\gamma^{2}}\left((4\omega^{2}_{0}-\gamma^{2})(1-e^{(-\gamma
    t)}cos(2\omega_{0}t))-4\omega_{0}\gamma e^{(-\gamma
    t)}sin(2\omega_{0}t)\right)\right] \\
  \tilde{\alpha}^{I} &=& \frac{1}{4\omega^{2}_{0}+\gamma^{2}}\left[-2\omega_{0}t+\frac{1}{4\omega^{2}_{0}+\gamma^{2}}\left(4\omega_{0}\gamma(1-e^{(-\gamma t)}cos(2\omega_{0}t))+(4\omega^{2}_{0}-\gamma^{2})e^{(-\gamma t)}sin(2\omega_{0}t)\right)\right]
\end{eqnarray}
\end{widetext}
\par The exponential functions of superoperators in Eq.(\ref{e8}) could be disentangled in
form\cite{pur}
\begin{eqnarray}
  \hat{T}e^{\int_{0}^{t}dt(\varepsilon_{0}J_{0}+ \varepsilon_{+}J_{+}+\varepsilon_{-}J_{-})} &=& e^{j_{+}J_{+}}e^{j_{0}J_{0}}e^{j_{-}J_{-}} \\
  \hat{T}e^{\int_{0}^{t}dt(\nu_{0}K_{0}+\nu_{+}K_{+}+\nu_{-}K_{-})} &=& e^{k_{+}K_{+}}e^{k_{0}K_{0}}e^{k_{-}K_{-}}
\end{eqnarray}
where $j_{+}$, $j_{0}$, $j_{-}$ and $k_{+}$, $k_{0}$, $k_{-}$
satisfy the following equation
\begin{eqnarray}
  \dot{X}_{+} &=& \mu_{+}-\mu_{-}X_{+}^{2}+\mu_{0}X_{+} \\
  \dot{X}_{0} &=& \mu_{0}-2\mu_{-}X_{+}\\
  \dot{X}_{-} &=& \mu_{-}exp(X_{0})
\end{eqnarray}
$\mu=\varepsilon$ for $X=j$ and $\mu=\nu$ for $X=k$.

\par Using the results in Appendix, the exact solution of the master
equation Eq.(\ref{e7}) is obtained
\begin{equation}\label{e19}
     \rho_{a}(t) = e^{-\Gamma_{k}}\tilde{\rho}(t)
\end{equation}
\begin{equation}\label{e20}
    \tilde{\rho}(t)=\left(\begin{array}{cc}
                           l\rho^{11}_{a}(0)+ m\rho^{00}_{a}(0) & x\rho^{10}_{a}(0)+ y\rho^{01}_{a}(0) \\
                           q\rho^{01}_{a}(0)+ r\rho^{10}_{a}(0) &
                           n\rho^{00}_{a}(0)+ p\rho^{11}_{a}(0)
                         \end{array}
                         \right)
\end{equation}
\begin{eqnarray}
  l &=& e^{k_{0}/2}+e^{-k_{0}/2}k_{+}k_{-},\ m = e^{-k_{0}/2}k_{+} \\
  n &=& e^{-k_{0}/2},\ p= e^{-k_{0}/2}k_{-}\\
  q &=& e^{-j_{0}/2},\ r= e^{-j_{0}/2}j_{-} \\
  x &=& e^{j_{0}/2}+e^{-j_{0}/2}j_{+}j_{-},\ y=e^{-j_{0}/2}j_{+}
\end{eqnarray}

\subsection{ Concurrence}
\par In order to investigate the entanglement dynamics of the bipartite
system, we use Wootters concurrence\cite{woott}. For simplicity, we
assume the two subsystems having the same parameters. The
concurrence of the whole system could  be obtained\cite{ikr}
\begin{eqnarray}
  C_{\xi} &=& max \left\{0,c_{1},c_{2}\right\},(\xi=\Phi,\Psi) \\
  c_{1} &=& 2e^{-2\Gamma_{k}}(\sqrt{\rho_{23}\rho_{32}}-\sqrt{\rho_{11}\rho_{44}}) \nonumber\\
  c_{2} &=&
  2e^{-2\Gamma_{k}}(\sqrt{\rho_{14}\rho_{41}}-\sqrt{\rho_{22}\rho_{33}})\nonumber
\end{eqnarray}
corresponding to the initial states of $|\Phi\rangle =
\beta|01\rangle+\eta|10\rangle$ and $|\Psi\rangle =
\beta|00\rangle+\eta|11\rangle$, respectively. Where $\beta$ is real
and $0<\beta<1$, $\eta=|\eta|e^{i\varphi}$ and
$\beta^{2}+|\eta|^{2}=1$. The reduced joint density matrix of the
two atoms, in the standard product basis
$\mathfrak{B}=\{|1\rangle\equiv|11\rangle,
|2\rangle\equiv|10\rangle, |3\rangle\equiv|01\rangle,
|4\rangle\equiv|00\rangle \}$, could be written as\cite{bel}
\begin{equation}\label{e28}
    \rho^{AB}=e^{-2\Gamma_{k}}
    \left(\begin{array}{cccc}
            \rho_{11} & 0 & 0 & \rho_{14} \\
            0 & \rho_{22} & \rho_{23} & 0 \\
            0 & \rho_{32} & \rho_{33} & 0 \\
            \rho_{41} & 0 & 0 & \rho_{44}
          \end{array}
          \right)
\end{equation}
here the diagonal elements are
\begin{eqnarray*}
  \rho_{11} &=& l^{2}\rho_{11}(0)+lm\rho_{22}(0)+ml\rho_{33}(0)+m^{2}\rho_{44}(0) \\
  \rho_{22} &=& lp\rho_{11}(0)+lm\rho_{22}(0)+mp\rho_{33}(0)+mn\rho_{44}(0) \\
  \rho_{33} &=& lp\rho_{11}(0)+pm\rho_{22}(0)+nl\rho_{33}(0)+nm\rho_{44}(0) \\
  \rho_{44} &=& p^{2}\rho_{11}(0)+pn\rho_{22}(0)+np\rho_{33}(0)+n^{2}\rho_{44}(0)
\end{eqnarray*}
and the nondiagonal elements are
\begin{eqnarray*}
  \rho_{14} &=& x^{2}\rho_{14}(0)+xy\rho_{23}(0)+yx\rho_{32}(0)+y^{2}\rho_{41}(0) \\
  \rho_{23} &=& xr\rho_{14}(0)+xq\rho_{23}(0)+yr\rho_{32}(0)+yq\rho_{41}(0) \\
  \rho_{32} &=& rx\rho_{14}(0)+ry\rho_{23}(0)+qx\rho_{32}(0)+qy\rho_{41}(0) \\
  \rho_{41} &=& r^{2}\rho_{14}(0)+rq\rho_{23}(0)+qr\rho_{32}(0)+q^{2}\rho_{41}(0)
\end{eqnarray*}

 \section{\label{sec:num}Numerical results and discussion}
 \par In order to study the effects of non-Markovian reservoir
 on the decoherence, we assume $\lambda=10\gamma$ in Eq.(\ref{ee1}), which could be realized in a hight-Q cavity\cite{bel}.
 \par
\begin{figure}
  \includegraphics{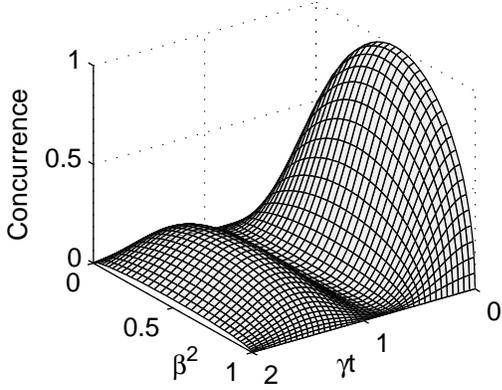}\\
  \caption{Concurrence $C_{\Phi}$ as a function of $\gamma t$ and $\beta^{2}$ with $\lambda=10\gamma$ for RWA model.}\label{F1}
\end{figure}

 First, we focus on the decoherence of
 two qubits with an initial state $|\Phi\rangle$. For the RWA model
in Ref.\cite{bel}, Fig.\ref{F1} shows that the concurrence
periodically vanishes with a damping of its revival amplitude. For
the non-RWA model in this paper, the decoherence of the system were
categorized into three cases
\par
\begin{figure}
  \includegraphics{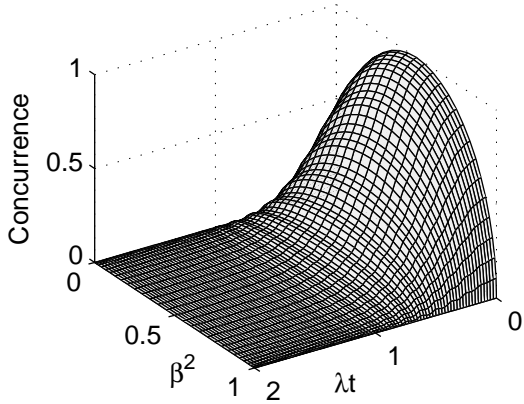}\\
  \caption{Concurrence $C_{\Phi}$ as a function of $\gamma t$ and $\beta^{2}$ with $\omega_{0}=10\lambda$, $\lambda=10\gamma$.}\label{F2}
\end{figure}

 (A) $\omega_{0}>\lambda, \omega_{0}\gg\gamma$. Fig.\ref{F2} reveals that
 the concurrence $C_{\Phi}$ decreases
exponentially to zero with very small amplitude oscillation and the
decoherence time is about $1/\gamma$, which corresponds to the
correlation time of $\alpha_{1}$ coming from the rotating-wave
terms. As for the correlation function $\alpha_{2}$, it has no
memory effect because it's value averages to zero on time if
$\gamma/\omega_{0}\ll1$. Compared with Fig.\ref{F1} for RWA model,
there is no revival of entanglement for non-RWA model, which
exhibits that the counter-rotating wave terms could not be neglected
in Hamiltonian because the contribution of counter-rotating terms in
master equation is in the order of
$\lambda\gamma^{2}/\omega_{0}^{2}$ which influence the decoherence
behavior in long time scale. This characteristic will hold on for
more weaker coupling constant.

\par
\begin{figure}
  \includegraphics{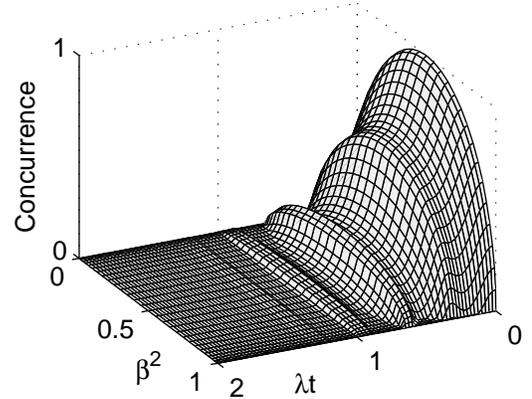}\\
  \caption{Concurrence $C_{\Phi}$ as a function of $\gamma t$ and $\beta^{2}$ with $\omega_{0}=\lambda=10\gamma$.}\label{F3}
\end{figure}

 (B) $\omega_{0}=\lambda>\gamma$. From Fig.\ref{F3}, we
could find that the concurrence $C_{\Phi}$ first decreases to a
finite value and maintain it for a period of time, then periodically
vanishes with a damping of its revival amplitude, like that for RWA
model. This is the result of coaction of rotating wave process and
counter-rotating process of atom with a non-Markovian reservoir
because the contribution of counter-rotating wave process to the
system is bigger than the former case and its memory effect is
nozero.

\par
\begin{figure}
  \includegraphics{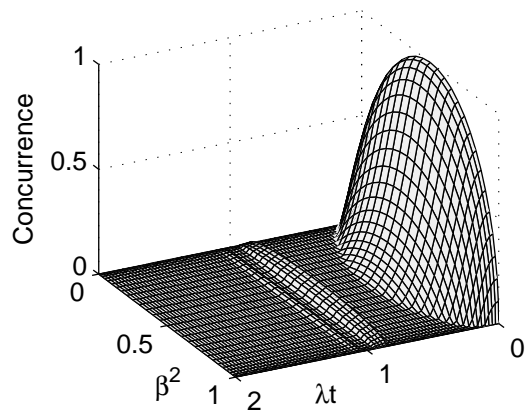}\\
  \caption{Concurrence $C_{\Phi}$ as a function of $\gamma t$ and $\beta^{2}$ with $\lambda=10\gamma$, $\omega_{0}=3\gamma$.}\label{F4}
\end{figure}

 (C) $\omega_{0}=3\gamma<\lambda$. Fig.\ref{F4}
exhibits that the concurrence $C_{\Phi}$ decreases to zero, then
periodically vanishes with a damping of its revival amplitude, which
resulted from the strong interaction between atom and the
non-Markovian reservoir throuth virtual photon process.

\par From the discussion above, we find that the decoherence behavior of $C_{\Phi}$
is symmetric with $\beta^{2}$ because of the symmetry of initial
state $|\Phi\rangle$. And the revival of entanglement could be found
only with big ratio of coupling constant to the atom transition
frequency. For no-RWA model, the decoherence behaviors are richer
than that for RWA model.

\par
\begin{figure}
  \includegraphics{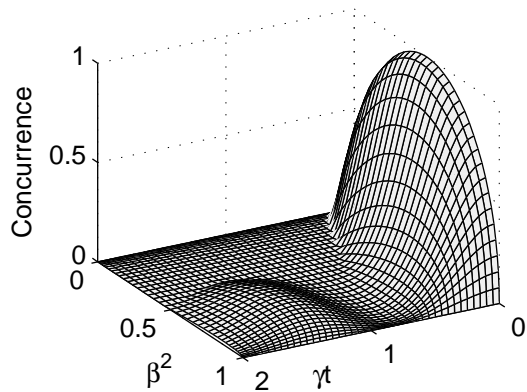}\\
  \caption{Concurrence $C_{\Psi}$ as a function of $\gamma t$ and $\beta^{2}$ with $\lambda=10\gamma$ for RWA model.}\label{F5}
\end{figure}

 Then, we focus on the decoherence of
 two qubits with initial state of $|\Psi\rangle$. For the RWA model
in Ref.\cite{bel}, Fig.\ref{F5} shows that the entanglement
represented by $C_{\Psi}$ has a similar behavior to $C_{\Phi}$ for
$\beta^{2}\geq1/2$ in Fig.\ref{F1}. In contrast, for
 $\beta^{2}<1/2$, there is ESD because $C_{\Psi}$ vanishes permanently after a finite
time, similar to the Markovian case\cite{yu2}. Second, revival of
entanglement appears after periods of times when disentanglement is
complete. For the non-RWA model, the decoherence of the system were
also categorized into three cases
\par
\begin{figure}
  \includegraphics{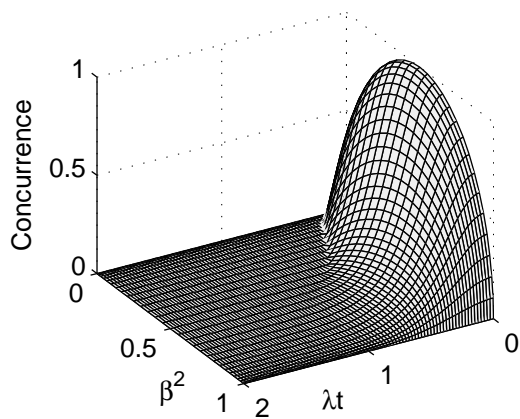}\\
  \caption{Concurrence $C_{\Psi}$ as a function of $\gamma t$ and $\beta^{2}$ with $\omega_{0}=10\lambda$, $\lambda=10\gamma$.}\label{F6}
\end{figure}

 (A) $\omega_{0}>\lambda, \omega_{0}\gg\gamma$. Fig.\ref{F6} reveals that
 the concurrence $C_{\Psi}$ decreases
exponentially to zero for $\beta^{2}\geq1/2$, while $C_{\Psi}$
vanishes permanently after a finite time for $\beta^{2}<1/2$.

\par
\begin{figure}
  \includegraphics{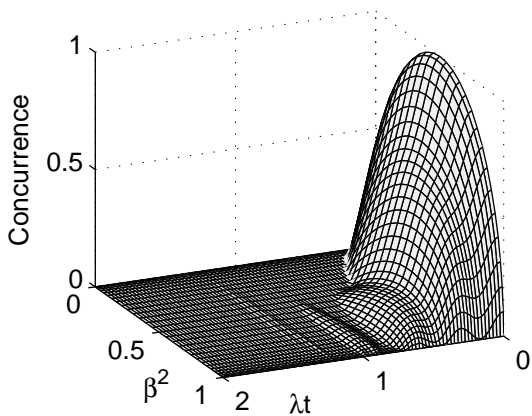}\\
  \caption{Concurrence $C_{\Psi}$ as a function of $\gamma t$ and $\beta^{2}$ with $\omega_{0}=\lambda=10\gamma$.}\label{F7}
\end{figure}

 (B) $\omega_{0}=\lambda>\gamma$. Fig.\ref{F7} shows that the entanglement
represented by $C_{\Psi}$ has a similar behavior to $C_{\Phi}$ for
$\beta^{2}\geq1/2$ in Fig.\ref{F3}. In contrast, for
 $\beta^{2}<1/2$, the concurrence first decreases to zero and maintain it for a long period of time,
 then revives with very small amplitude before vanishes permanently.

\par (C) $\omega_{0}=3\gamma<\lambda$. The evolution dynamics of concurrence $C_{\Psi}$ is almost
the same as that in Fig.\ref{F4}. Unlike the two cases above, the
evolution behavior of concurrence $C_{\Psi}$ becomes symmetric, like
that of $C_{\Phi}$, because of the strong interaction of atom with
reservoir through the emission and absorption of virtual photon.

\par The above characteristics of $C_{\Psi}$ also verify that the revival of entanglement could be found
only with big ratio of coupling constant to the atom transition
frequency.
\section{\label{sec:clu}Conclusion}
\par The reduced non-perturbative non-Markovian quantum master
equation of atom in non-Markovian reservoir has been derived and its
exact solution is obtained by algebraic approach of Lie
superoperator. The decoherence of two initially entangled atoms,
coupled with two vacuum cavities separately, has been discussed.
\par The results show that the decoherence behavior of two qubits in
a non-markovian reservoir is dependent on the ratio of the coupling
strength to atomic frquency. First, with the increasing of coupling
 strength, the decoherence behavior becomes more and more
complicated, and there is revival of entanglement after a period of
time of disentanglement, due to the strong interaction and the
memory effect of the non-markovian reservoir. Second, the strong
coupling and the counter-rotating wave interaction could make the
decoherence of even parity Bell entanglement state become
symmetrical. Third, with no-RWA model, one could find much more
decoherence behaviors than that of RWA model under different
conditions of system parameters. The case for finite temperature
could also be obtained by the method in this paper and will be
published elsewhere.

\begin{acknowledgments}
This work was supported by the National Natural Science Foundation
of China Grants No.60578055, the State Key Program for Basic
Research of China under Grant No. 2007CB925204 and No.2007CB307001.
\end{acknowledgments}
\appendix
\section{}
\par Defining two superoperators as
\begin{eqnarray}
  J &\equiv&  e^{j_{+}J_{+}}e^{j_{0}J_{0}}e^{j_{-}J_{-}} \\
  K &\equiv&  e^{k_{+}K_{+}}e^{k_{0}K_{0}}e^{k_{-}K_{-}}
\end{eqnarray}
and using the following relations
\begin{eqnarray}
  e^{j_{+}J_{+}}\rho &=& \rho+j_{+}\sigma_{+}\rho\sigma_{+} \\
  e^{j_{-}J_{-}}\rho &=& \rho+j_{-}\sigma_{-}\rho\sigma_{-}\\
  e^{k_{+}K_{+}}\rho &=& \rho+k_{+}\sigma_{+}\rho\sigma_{-} \\
  e^{k_{-}K_{-}}\rho &=& \rho+k_{-}\sigma_{-}\rho\sigma_{+}
\end{eqnarray}
and
\begin{widetext}
\begin{eqnarray}
  e^{j_{0}J_{0}}\rho &=& (ch\frac{j_{0}}{4}+\sigma_{z}sh\frac{j_{0}}{4})\rho(ch\frac{j_{0}}{4}-\sigma_{z}sh\frac{j_{0}}{4}) \\
  e^{k_{0}K_{0}}\rho &=&
  e^{\frac{k_{0}}{2}}[1+(e^{\frac{k_{0}}{2}}-1)\sigma_{+}\sigma_{-}]\rho[1+(e^{\frac{k_{0}}{2}}-1)\sigma_{+}\sigma_{-}],
\end{eqnarray}
\end{widetext}
one could get
\begin{widetext}
\begin{eqnarray}
    J \left(
    \begin{array}{cc}
       \rho_{11} & \rho_{10}  \\
       \rho_{01} & \rho_{00}
     \end{array}\right)&=& \left(
     \begin{array}{cc}
       \rho_{11} & (e^{j_{0}/2}+e^{-j_{0}/2}j_{+}j_{-})\rho_{10}+ e^{-j_{0}/2}j_{+}\rho_{01} \\
       e^{-j_{0}/2}\rho_{01}+ e^{-j_{0}/2}j_{-}\rho_{10} & \rho_{00}
     \end{array}
     \right) \\
   K \left(
    \begin{array}{cc}
       \rho_{11} & \rho_{10}  \\
       \rho_{01} & \rho_{00}
     \end{array}\right)&=& \left(
     \begin{array}{cc}
       (e^{k_{0}/2}+e^{-k_{0}/2}k_{+}k_{-})\rho_{11}+ e^{-k_{0}/2}k_{+}\rho_{00} & \rho_{10} \\
      \rho_{01} &  e^{-k_{0}/2}\rho_{00}+ e^{-k_{0}/2}k_{-}\rho_{11}
     \end{array}
     \right)
\end{eqnarray}
\end{widetext}

\newpage 
\bibliography{apssamp}

\end{document}